\documentclass[10pt,conference]{IEEEtran}
\IEEEoverridecommandlockouts
\usepackage{braket}
\usepackage{cite}
\usepackage{amsmath,amssymb,amsfonts}
\usepackage[numbers,sort&compress]{natbib}
\usepackage{algorithmic}
\usepackage{graphicx}
\usepackage{textcomp}
\usepackage{xcolor}
\usepackage{slashed}
\usepackage{hyperref}
\hypersetup{
    colorlinks=true,
    linkcolor=blue,
    filecolor=magenta,      
    urlcolor=blue,
    pdfpagemode=FullScreen,
    }

\usepackage{multirow}
\usepackage{tabularx}
\usepackage{caption}
\def\BibTeX{{\rm B\kern-.05em{\sc i\kern-.025em b}\kern-.08em
    T\kern-.1667em\lower.7ex\hbox{E}\kern-.125emX}}

\begin{document}

\title{On the importance of scalability and resource estimation of quantum algorithms for domain sciences}

\author{\IEEEauthorblockN{Vincent R. Pascuzzi\thanks{Corresponding author: \href{mailto:pascuzzi@bnl.gov}{pascuzzi@bnl.gov}}}
\IEEEauthorblockA{\textit{Computational Science Initiative} \\
\textit{Broohaven National Laboratory}\\
Upton, USA \\
0000-0003-3167-8773}
 \and
 \IEEEauthorblockN{Ning Bao}
 \IEEEauthorblockA{\textit{Computational Science Initiative} \\
 \textit{Broohaven National Laboratory}\\
 Upton, USA \\
 0000-0002-3296-1039}
 
  \and
 \IEEEauthorblockN{Ang Li}
 \IEEEauthorblockA{\textit{HPC Group, ACMD, PCSD} \\
 \textit{Pacific Northwest National Laboratory}\\
 Richland, WA, USA \\
 0000-0003-3734-9137}
}

\maketitle

\begin{abstract}
The quantum information science community has seen a surge in new algorithmic
developments across scientific domains.
These developments have demonstrated polynomial or better improvements in computational
and space complexity, incentivizing further research in the field.
However, despite recent progress, many works fail to provide quantitative
estimates on algorithmic scalability or quantum resources required
---\textit{e.g.}, number of logical qubits, error thresholds, \textit{etc.}---to
realize the highly sought ``quantum advantage.''
In this paper, we discuss several quantum algorithms and motivate the importance
of such estimates.
By example and under simple scaling assumptions, we approximate the capabilities
needed of a future quantum device for a high energy physics simulation algorithm
to achieve superiority over its classical analog.
We assert that a standard candle is necessary for claims of quantum advantage.
\end{abstract}


\section{Introduction}
Quantum computation and information are interdisciplinary marvels born
out of the convergence of arguably the two most transformational sciences in
human history: Physics and Computer Science.
By Dennard scaling, and the laws of Moore and Amdahl, we are nearing
the physical limitations of classical computing.
Quantum information processing promises exponential gains in terms of
computational and space complexity for special classes of problems across
scientific domains and industrial applications.
Exploiting quantum mechanical phenomena---such as superposition, interference,
and entanglement---holds great promise for constructing next-generation devices
which will complement traditional von Neumann architectures, enabling
researchers to solve currently intractable problems.
The development of quantum algorithms, however, requires tremendous ingenuity
to leverage the advantages quantum physics offers.
Although many such algorithms exist which show a quantum
advantage~\cite{10.1137/S0097539795293172, grover1996fast, coppersmith2002approximate}, the lack of error-correction or fault-tolerance supersedes
algorithms' quantum hallmarks, rendering them either impractical on
present day systems or requiring simplification to the point that a classical
computer could also perform the task with tractable resource consumption.
As such, there exists a serious need for researchers of quantum algorithms to
heed practical benefits and considerations when claiming quantum superiority, particularly those primarily motivated by domain science application.

A central area with which the author is familiar is the area of quantum simulation.
Here, the task is to use a quantum computing device to simulate the dynamics of
a quantum mechanical or field theoretic system of physical interest.
While this is a problem that is of clear physical interest to many different
subfields of physics and chemistry, it is also clear that the reach of direct
quantum simulation of these areas is severely limited in the current area of
weak and relatively unreliable quantum computers, often only able to access
small lattice sizes which are classically simulable, or requiring dramatic
simplification to the point of straining physical applicability.
While doing some simulations of this type is useful for proof-of-principle,
doing large numbers of such simulations is perhaps not, without a clear and
attainable scientific goal.

Something that would be useful in this regard, however, is far more detailed
benchmarking of quantum simulation algorithms for domain science in terms of
resource requirements in, \textit{e.g.}, gate depth or qubit number.
Such estimates would much better enable scientists to determine when quantum
computing technology has sufficiently advanced in order to study scientifically
relevant questions using quantum simulation.
Such benchmarking should be done with current qubit and gate quality in mind,
and should be updated with the progression of quantum
computing technology.
In this way, it will by construction provide an upper bound on the number of
years/advances required in order to run a scientifically
useful quantum algorithm that grows ever closer with both the passing of time
and the development of the technology.

\section{Quantum Algorithms}
\subsection{Quantum Fourier Transform}
\begin{figure}[!t]
\centering
\includegraphics[width=1\columnwidth]{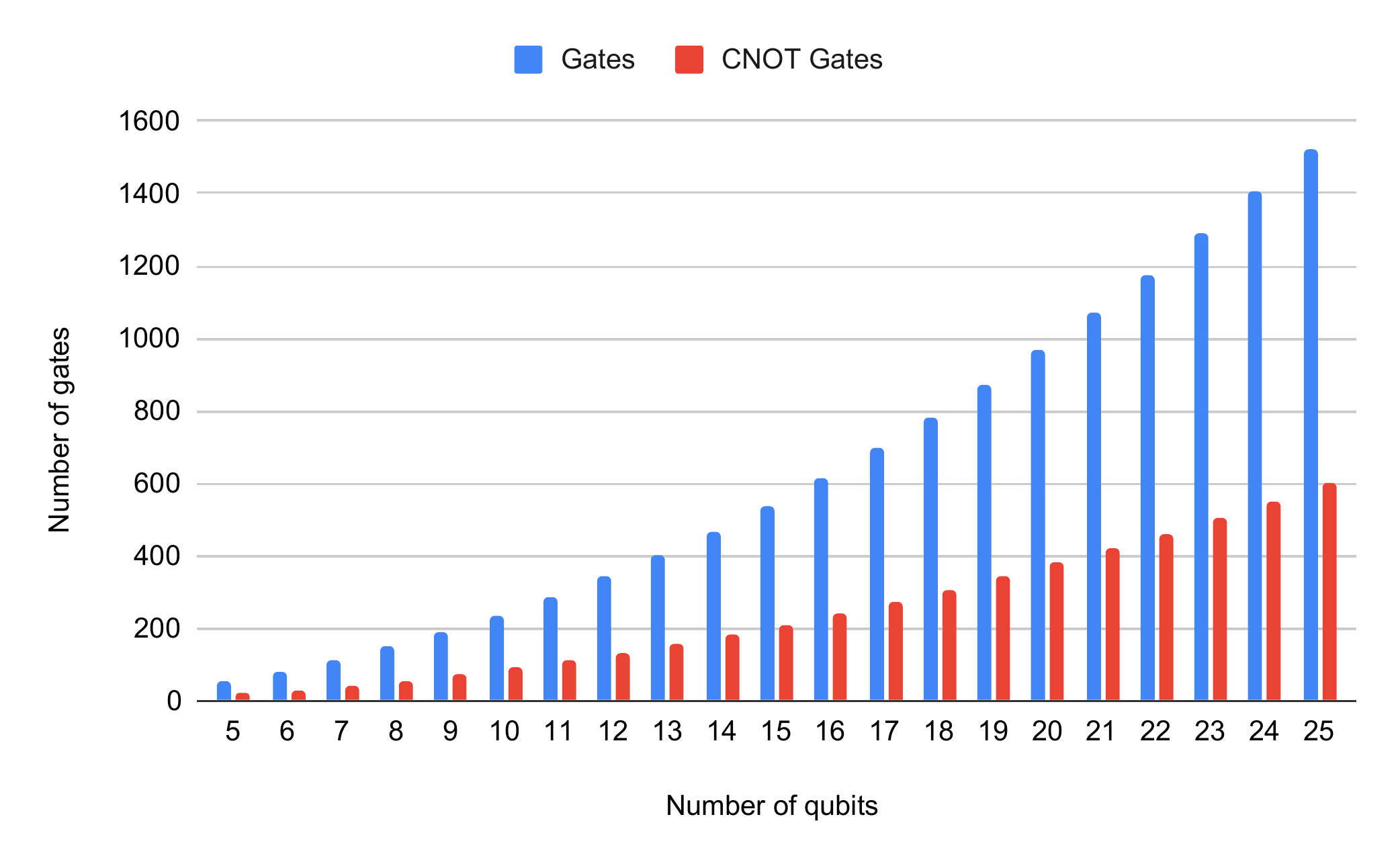} 
\caption{The scaling of gates w.r.t. qubits for QFT.}
\label{fig:qft}
\end{figure}

The Quantum Fourier Transform (QFT) \cite{coppersmith2002approximate} (and its inverse) applies (inverse) Fourier transformation to the wave function amplitudes. It is a linear transformation over the states of qubits, which is the quantum analogue of the discrete (inverse) Fourier transform. QFT is a basic component of many well-known quantum algorithms, including Shor's algorithm \cite{10.1137/S0097539795293172}, quantum phase estimation \cite{dorner2009optimal}, hidden subgroup problem \cite{mosca1998hidden}, etc. Fig.~\ref{fig:qft} shows the scaling of the number of gates and particularly CNOT gates with respect to the number of qubits for QFT from 5 to 25. As can be seen, the number of gates and CNOT gates scale exponentially with increased qubits. 

\subsection{Grover's Algorithm}

Grover's algorithm \cite{grover1996fast} is a quantum algorithm for searching in a database, offering quadratic speedup over classical searching methods. It takes a black-box oracle realizing the function: $f(x)=1$ if $x=y$, $f(x)=0$ if  $x\ne y$, and find $y$ within a randomly ordered sequence of $N$ items using $O(\sqrt{N})$ operations and $O(N\log{N})$ gates with a probability of $p \ge 2/3$. Grover's algorithm is appealing in that it determines with high probability the unique input given the output is pre-known. Grover's algorithm is one of the fundamental quantum algorithms given the importance of searching in data-driven sciences.

\subsection{Harrow-Hassidim-Lloyd}

Harrow-Hassidim-Lloyd algorithm (HHL) is one of the fundamental algorithms expected to offer substantial speedup over the classical one and of great interest in many domains. HHL algorithm is a quantum algorithm for linear systems of equations. Given a sparse system with a low condition number $\kappa$, for a scalar measurement of the solution vector, the HHL algorithm can demonstrate a runtime of $O(\log(N)\kappa^2)$ where $N$ is the number of variables, comparing to the best known classical approach of $O(N\kappa)$.

\subsection{Hybrid Quantum-Classical Algorithms}

The variational quantum algorithm is to optimize a parameterized quantum circuit ansatz applied to some initial state for minimizing a cost function defined according to the output state \cite{mcclean2016theory, jones2019variational}. When applied in quantum simulation, the goal of the algorithm is often to prepare ground states. The cost function is often the expectation value of a Hamiltonian. If the initial state is $\ket{\psi}$, the Hamiltonian is $H$, the ansatz is $U(\vec{\theta})$ where $\vec{\theta}$ is the variational parameter, then the cost function is $E(\vec \theta) = \bra{\psi} U^\dagger(\vec{\theta}) H U(\vec{\theta}) \ket{\psi}$.

The quantum approximate optimization algorithm (QAOA) \cite{farhi2014quantum} is another variational quantum algorithm that is particularly interesting, given its ability to tolerant certain degree of noise through the unique quantum-classic hybrid algorithm design. QAOA is designed to solve combinatorial optimization problems, such as in graph analytics. In QAOA, a quantum subroutine is embedded in a classical search loop. The quantum state is prepared according to a number of variational parameters. These parameters are adjusted based on the measurement result per iteration.

\section{Considerations for claims of ``quantum advantage''}
\label{sec:considerations}
Quantum computing is inherently more `difficult' than that of classical;
not solely due to the different way of thinking about computation but also
due to the complexity of quantum systems and the precise control necessary
for manipulating states.
As quantum computing systems become larger (qubit count) and increase
connectivity (highly entangled), many of today's algorithms which presently
simplify problems can be employed in more real-life scenarios.
However, the big questions here are `if' and `when' such systems will be
realized.
The full potential of practical quantum computing has yet to be realized,
and for the sake of this fascinating science and moving beyond the hype,
the community must stress the importance of  applications on near-future
devices.
Therefore, we assert that scalability and resource estimation should be
just as important in the design of quantum algorithms as the algorithms
themselves if we wish to build upon the momentum in this field.

\subsection{Scalability}
\label{ssec:scalability}
Scalability studies are needed to determine whether or not a quantum
algorithm maintains its advantage over its classical analog (if one
exists).
Determining scalability of a quantum algorithm, however, is somewhat
more involved than in classical computing;
one must consider scaling not only as additional computational resources
(qubits) are added to solve a problem, but also requirements such as
connectivity for highly entangled systems, and noise and error
thresholds---both SPAM and those which accrue during execution---which
dictate how deep a circuit can reliably execute.
For practical reasons, scaling may not be interesting when going from a
weakly connected system of $100$ qubits to another weakly connected
system of $100,000$ qubits;
rather, from a weakly connected $100$ qubit system to a strongly
connected system of $1,000$ qubits.

As such, there is no single figure of merit for describing a scalable
quantum algorithm;
algorithm developers need to either consider a collective set of
criteria or design algorithms with significant reliance on one over the
others.
These and other considerations may soon be necessary to demonstrate
further that quantum algorithms are more than merely pen-and-paper
demonstrations of solutions to real problems, but can and will be used
to advance science.

\subsection{Resource Estimation}
\label{ssec:res-est}
Beyond scalability is the need for estimating the computational
resources required to execute a quantum algorithm.
The estimates can pertain to near-term requirements or longer-term;
regardless, while estimates are by definition approximations or
judgements based (hopefully) on experience or trends, they give some
level of assurance an algorithm will be implemented and executed to
solve a full problem---as opposed to an overly simplified one---by
some device eventually.
Whether `eventually' may be within the next year, five years, decade
or later will determine the usefulness of a given quantum algorithm,
as one needs to also take into account the rapid technological
advances in classical computing.
For example, the GNFS and Shor's algorithms described earlier:
despite Shor's exponential speed-up, the largest number factorizable
on today's quantum systems is four bits (\textit{cf.} any
$10^{100}$ bit numbers, apart from prime powers, that GNFS can
factor).

Obviously exponential speed-ups of quantum algorithms over their
classical counterparts have been devised, and have also shown to be
scalable.
However, will they actually ever be able to run and empirically show
the quantum advantage?
Part of the answer to this question can be at least partially answered
through na\"{i}ve application of a quantum version of Moore's Law: the
number of qubits in a system doubles every two years.
The remaining considerations include: the evolution of qubit coherence
times (T1 and T2);
noise, fidelity and duration of physical gate sets
and thresholds thereof;
as well as other device-level capabilities, \textit{e.g.}, mid-circuit
measurements.
A concrete example is the well-known algorithm for simulating general
quantum field theories from Jordan, Lee and Preskill~\cite{jlp}.
This marvelous algorithm considers the fundamental stages in quantum
simulation: (1) state preparation (vacuum state, $\ket{0\ldots 0}$); (2)
adiabatic evolution (from non-interacting theory to interacting); (3)
Trotterization (time-evolution of interacting theory); and (4)
performing a measurement.
A back-of-the-envelope estimate of the number of error-corrected
(logical) qubits to needed to simulate a single process (\textit{e.g.},
$p\bar{p} \to \pi\pi\pi\pi$) is $O(10^7)$.
Given the current trends, we can expect perhaps $10^5$--$10^6$ physical
qubits within the next decade;
this does not consider connectivity or noise.
So the question is: what real advantage will there be over classical
methods in $\le 10$ years from now?

The take-home message here is straightforward: it is essential for
researchers to give some level of promise that their algorithm will
not only one day be demonstrated on real hardware, as well as what
advantage can be expected given some rough idea as to the classical
computational resources available to solve the same problem (either
via classical means or classically simulating quantum computations).

\subsection{Applicability and Relevance of the Algorithm}
\label{ssec:phys-rel}
This is an area where for example the HHL algorithm struggles. While it is true that the solution to the linear system is encoded in the coefficients of the quantum state, an exponential number of queries are needed to extract the full solution, thus negating the computational advantage. Only when a small number of queries of the quantum state are necessary in order to answer a problem of scientific relevance does this algorithm convey a genuine computational advantage over classical solvers.

Because many existent quantum algorithms are targeted at problems in mathematics and computer science that are interesting in large part (or solely) due to their difficulty, this question of physical relevance is unfortunately one that is ubiquitous in the assessment of quantum algorithms: even given an algorithm of known speed-up, worthy scalability characteristics, and reachable resource requirements both now and in the future, there is still a question to be asked of whether such an algorithm is useful for a reason beyond simply its difficulty.

\section{Case Study: A Quantum Algorithm for High Energy Physics Simulations}
As described in Sec.~\ref{ssec:phys-rel}, physically-motivated problems tend
to be of greater interest than those aiming to solve problems due primarily
to their inherent difficulty.
Keeping with this theme, we consider a recent quantum algorithm with
exponential speed-up in the context of high energy physics
(HEP)~\cite{PhysRevLett.126.062001};
specifically, a quantum final state algorithm
The reason for choosing this particular algorithm is four-fold:
(1) quantum simulation is a major driver in the community; (2)
its physical relevance; (3) it provides a meaningful scaling
analysis; and (4) scientific bias [the author(s) of this paper
come from a HEP background].

Simulations of physical phenomena are crucial in many aspects
of HEP---from conceptualizing to design of a new experiment,
to validation and operation.
Beyond this, simulations play a major role in many precision
measurement and searches for new physics.
High energy collisions, such those between protons at the
Large Hadron Collider, have to date been simulated
classically using perturbative (hard-scatter, shortest
distances), nonpertubative (hadronization, longest distances)
and Markov Chain Monte Carlo (MCMC; parton showers,
intermediate distances) techniques.
While these methods describe different length and energy
scales sufficiently well, physics analyses are becoming
increasingly more sensitive to quantum effects which are
not captured classically;
in particular, the efficiency of parton shower algorithms
stems from the fact that certain quantum effects are
neglected.
As more precise studies of final state radiation in high
energy collisions are performed, quantum effects---such as
interference---must be taken into account.

\subsection{Algorithm Overview}
\label{ssec:parton-motive}
The work from Nachman \textit{el al.} considered here
introduces a quantum algorithm targeting final state radiation
that gives, under certain conditions, exponential improvements
in both space and time.
The algorithm considers a simple quantum field theory with two
fermion fields interacting (non-zero couplings) with one scalar
boson.
It requires a total of six registers to hold information about
the particle state ($\ket{p}$) and number of each particle type
($\ket{n_\phi}, \ket{n_{f_a}}, \ket{n_{f_b}}$), whether
emission took place in a given step ($\ket{e}$) of the algorithm
and emission history ($\ket{h}$).
Evolution is performed via a sequence of four operations: (1)
$U_\text{count}$ is controlled on the particle state register and applies
the counts of each particle type to the corresponding count register; (2)
$U_e$ applies Sudakov factors to determine if an emission took place in
the step; (3) $U_h$ generates a superposition of all possible emissions
to select the particle that emitted;
and (4) $U_p$ which updates appropriately the particle register if an
emission occurred.
These operations are sandwiched between rotations to first change into
a diagonal basis and next to create interferences between equivalent final
states containing the same intermediate fermions.
For $N$ steps of the algorithm, up to $N$ particles can be emitted;
a collection of steps comprises an \textit{event}, and given a large
set of events one can then calculate physical observables of the
theory.

\subsection{Scaling}
\label{ssec:parton-scaling}
For brevity, we discuss here only the dominant space and time complexities
of the quantum parton shower algorithm.
For full details, we refer the interested reader to the paper's
supplemental material.

The two dominant registers in terms of the number of required qubits
are $\ket{p}$, requiring $3(N+n_I)$ qubits, and $\ket{h}$, 
requiring $N \lceil \log(N+n_I) \rceil$, where $N$ is the number of
steps and $n_I$ is the number of initial particles.
Note that states with different particle histories (\textit{i.e.},
different emitting particles and different intermediate fermions) do
not interfere with each other.
Therefore, $\ket{h}$ can be measured after $U_h$ is applied and can
then be reset to the state $\ket{0}$.
However, at the time the algorithm was developed, mid-circuit measurements
were not supported by IBM Quantum (IBM Q) devices and so the particle history
register could not be measured and reset in each step.
With this capability now available, one needs only
$\lceil \log (2n_I+1) \rceil$ number of qubits to store $\ket{h}$.

The depth (time) complexity with mid-circuit measurement is
$N {n_f}^2 \log n_f$, coming from the the history operator,
$U_h$, where $n_f$ is the number of fermions.
Comparing to the classical counterpart, with scaling $N 2^{n_f/2}$ that
is exponential in the number of fermions, the quantum parton shower
algorithm provides advantage when the number of emitted particles becomes
larger than 21.
This is a promising result, considering charged particle
multiplicities alone expected at the High Luminosity LHC will exceed
$O(100)$.
Moreover, the quantum algorithm is able to efficiently incorporate
important interference effects that are ignored in the classical
algorithm. \\

\noindent\textbf{Space Complexity}: $3(N+n_I)$ qubits \\
\textbf{Time Complexity}: $N {n_f}^2 \log n_f$ native gates

\subsection{Resource Estimates}
\label{ssec:parton-resource}
The quantum field theory considered is given by,
\begin{align}
\label{eqn:lagrangian}
    \mathcal{L} = &\bar{f_1} (i \slashed{\partial} + m_1) f_1
    + \bar{f_2} (i \slashed{\partial} + m_2) f_2 + (\partial_\mu \phi)^2 \nonumber \\
    &+ g_1 f_1 \bar{f_1} \phi + g_2 f_2 \bar{f_2} \phi
    + g_{12} (\bar{f_1} f_2 + \bar{f_2} f_1) \phi,
\end{align}
where $f_1$ and $f_2$ are fermions and $\phi$ a scalar boson.
Despite its simplicity, the resources necessary to simulate fully this theory with
non-zero couplings $g_i$, $g_{ij}$ are out of reach on current NISQ devices.
To realize the quantum parton shower algorithm on a real quantum system, the problem
was further simplified by excluding $\phi \to f \bar{f}$ splitting, ignoring the
running coupling and having a single fermion in the initial state.
The noisy quantum systems set further constraints, limiting the algorithm to four
steps.
The weak connectivity of current quantum devices further complicates matters, as
expensive SWAP gates are necessary to operate on qubits without physical connections.
Whereas the full theory in Eqn.~\ref{eqn:lagrangian} would require $O(10^4)$ gates
for four steps, the (very) simplified version requires 53.
However, despite its simplicity, the quantum parton shower algorithm manages to
capture important interference effects that are ignored in classical MCMC.

The quantum parton shower algorithm was demonstrated on the now-retired IBM Q
Johannesburg system using five (Q7, Q10--Q13) of its 20 qubits.
The qubit and gate properties of this device are given in Tables~\ref{tab:qubits} and~\ref{tab:gates} were retrieved from the backend in August 2020.
\begin{table}[t]
\centering
    \begin{tabularx}{0.98\columnwidth}{c c c c c c}
    \hline \hline
 Qubit & T1 [$\mu$s] & T2 [$\mu$s] & Readout Error & Pr($0\vert1$) & Pr($1\vert0$) \\
 \hline
  7  & 51.075 & 14.306 & 0.067 & 0.072 & 0.063 \\
  10 & 61.980 & 12.312 & 0.970 & 0.076 & 0.117 \\
  11 & 39.295 & 8.293  & 0.072 & 0.071 & 0.074 \\
  12 & 71.769 & 13.577 & 0.101 & 0.108 & 0.094 \\
  13 & 82.397 & 17.041 & 0.088 & 0.088 & 0.089 \\
  \hline
  Avg.& 61.303 & 13.106 & 0.418 & 0.083 & 0.087 \\ 
\hline \hline
    \end{tabularx}
    \caption{IBM Q Johannesburg qubit properties.
    Data retrieved August 2020.}
    \label{tab:qubits}
\end{table}
\begin{table}[t]
\centering
    \begin{tabularx}{0.83\columnwidth}{c c c}
    \hline \hline
 Gate & Error & Duration [ns] \\
 \hline
  $U3\_7, C_X7\_12$   & $0.0019, 0.0264$ & $71.111, 490.6667$ \\
  $U3\_10, C_X10\_11$ & $0.0022, 0.0184$ & $71.111, 298.6667$ \\
  $U3\_11, C_X11\_12$ & $0.0015, 0.0265$ & $71.111, 483.5556$ \\
  $U3\_12, C_X12\_13$ & $0.0015, 0.0167$ & $71.111, 348.4444$ \\
  $U3\_13$            & $0.0014$         & $71.111$           \\
  \hline
  Avg.                & $0.0017, 0.0209$ & $71.111, 393.9556$ \\ 
\hline \hline
    \end{tabularx}
    \caption{IBM Q Johannesburg gate properties.
    $C_X$ gates are assumed symmetric ($C_{X_{ij}} = C_{X_{ji}}$) so redundant
    entries are omitted.
    Data retrieved August 2020.}
    \label{tab:gates}
\end{table}
Note the $U3 = U3(\pi, 0, \pi)$ represent Pauli $X$ gates, and we take $C_X$ gates
to be symmetric with respect to error and duration (\textit{i.e.},
$C_{X_{ij}} = C_{X_{ji}}$).
The state-of-the-art IBM Q Montreal device comprises 27 qubits an average T1 of
roughly 130~$\mu$s, T2 of 97~$\mu$s, and mis-measurement probabilities [Pr($0\vert1$), Pr($1\vert0$)] are cut by more than half compare to Johannesburg.
Based on report gate properties, $C_X$ errors have also improved by 50\% or more.
A na\"{i}ve scaling of resources thus yields roughly double overall performance
every two years, perhaps making possible circuit depths (additional steps)
permitting $O(10)$ particle multiplicities within the next five
years but otherwise the same limitations mentioned earlier.
It is therefore inconceivable a quantum advantage will be realized within the
next decade.
Notwithstanding, the quantum parton shower algorithm is an examplar demonstration of
innovation that has potential to be leveraged as quantum computing devices become
more advanced.

\section{Conclusions}
\label{ssec:parton-conclusion}
In this paper, we discussed and emphasized the need for scalability and resource
estimation studies in the context of quantum algorithm design.
There is without doubt great excitement in the field, and high hopes the
second quantum revolution will be able to provide greater insight into complex
phenomena.
The tremendous interest among domain scientists in quantum information and computing
is well-deserved, but the community best heed caution as the mutual interest from
funding agencies may very well subside without practical near-term applications and
results, leaving a small number of experts to continue their research in a
field become stagnant.

The development of quantum algorithms holds prodigious potential for impact in domain
science and beyond. In order to precisely quantify the impact in the near,
intermediate, and future terms, we advocate for better benchmarks are needed for
novel quantum algorithms, particularly those developed with domain science in mind by
scientists without traditional theoretical computer science training.
Such benchmarking will allow for clearer understanding for which algorithms can
provide lasting impact for domain science.

This paper was not meant to be pessimistic, rather, to fuel further innovations
from the community so the science can continue to flourish.

\section*{Acknowledgment}
NB and VRP are both supported by Brookhaven National Laboratory.
This work was not funded directly---it materialized out of joy and passion.

\bibliographystyle{unsrtnat}
\bibliography{main}

\end{document}